# The exploration of the unknown

**K. I. Kellermann**[1]
*National Radio Astronomy Observatory*
*Charlottesville, VA, U.S.A.*
*E-mail:* kkellerm@nrao.edu

J. M. Cordes
*Cornell University*
*Ithaca, NY*
*E-mail:* cordes@astro.cornell.edu

R.D. Ekers
*CSIRO*
*Sydney, Australia*
*E-mail:* Ron.Ekers@csiro.au

J. Lazio
*Naval Research Lab*
*Washington, D.C., USA*
*E-mail:* lazio@nrl.navy.mil

Peter Wilkinson
*Jodrell Bank Center for Astrophsyics*
*Manchester, UK*
*E-mail:* pnw@jb.man.ac.uk

[1] Speaker

## Summary

The discovery of cosmic radio emission by Karl Jansky in the course of searching for the source of interference to telephone communications and the instrumental advances which followed, have led to a series of new paradigm changing astronomical discoveries. These include the non-thermal emission from stars and galaxies, electrical storms on the Sun and Jupiter, radio galaxies, AGN, quasars and black holes, pulsars and neutron stars, the CMB, interstellar molecules and giant molecular clouds; the anomalous rotation of Venus and Mercury, cosmic masers, extra-solar planets, precise tests of gravitational bending, gravitational lensing, the first experimental evidence for gravitational radiation, and the first observational evidence for cosmic evolution. These discoveries, which to a large extent define much of modern astrophysical research, have resulted in eight Nobel Prize winners. They were the result of the right people being in the right place at the right time using powerful new instruments, which in many cases they had designed and built. They were not as the result of trying to test any particular theoretical model or trying to answer previously posed questions, but they opened up whole new areas of exploration and discovery. Rather many important discoveries came from military or communications research; others while looking for something else; and yet others from just looking.

Traditionally, the designers of big telescopes invariably did not predict what the telescopes would ultimately be "known for," so we may anticipate that the place in history of the next generation of telescopes will not likely be found in the science case created to justify their construction, but in the unexpected new phenomena, new theories, and new ideas which will emerge from these discoveries.

It is important that those who are in a position to filter research ideas, either as grant or observing time referees, as managers of facilities, or as mentors to young scientists, not dismiss as "butterfly collecting," investigations which explore new areas of phase space without having predefined the result they are looking for. Progress must also allow for new discoveries, as well as for the explanation of old discoveries. New telescopes need to be designed with the flexibility to make new discoveries which will invariably raise new questions and new problems.

## 1. Introduction

Astrophysics is an observational science. Unlike most scientists, astronomers are not able to do experiments, but can only observe the sky with open "eyes." We are dependent on a variety of emission processes complicated by a variety of absorption mechanisms, but we try to observe, and we try to understand what we see. Since Galileo's pioneering observations of sunspots, craters on the moon, the phases of Venus, the satellites of Jupiter, and the rings of Saturn,[1] astronomers, using instruments of ever increasing sophistication (and cost) have made a series of remarkable discoveries, only a few of which have resulted from attempts to test theoretical predictions. The existence of other galaxies, novae, and supernovae, dark matter, and dark energy were all first recognized from their observational discovery. Arguably, the most remarkable changes in the astronomical landscape began only in the 20th century, many as a result of observations made at radio wavelengths, as well as others that were unanticipated.

For more than three centuries after Galileo's discoveries, astronomical observations were confined to the narrow optical window closely corresponding to the sensitivity of the human eye. With the extension, some 75 years ago, to the broad radio spectrum covering more than eight decades of wavelength and later the expansion to space based facilities to access the IR, UV, X and γ-ray parts of the electromagnetic spectrum astronomers have access to the entire electromagnetic spectrum from below 1 MHz to photon energies above one TeV. Modern astrophysical research currently deals with questions and phenomena undreamed of even a few decades ago. While it is important to delineate the questions and problems to be addressed by the next generation of astronomical facilities which will lead to a better understanding of these recently discovered phenomena, it is equally important to design the new facilities to optimize their potential for the discoveries which will raise new questions and new problems.

## 2. The Lessons of History: Astronomical Discoveries

Planning for the unexpected can be challenging, but there is perhaps something to be learned from understanding the circumstances leading to past discoveries and how they have changed our perception of the large scale properties of the Universe and the nature of its constituents.

Because radio astronomy was the first of the new astronomies to explore the rich region beyond the classical optical/NIR spectrum, observations at radio wavelengths have been particularly rewarding in disclosing new, previously unknown, cosmic phenomena. Later, space borne facilities opened up the rich high energy sky. Most of these discoveries serendipitously resulted from investigations targeted at other astronomical problems, but some were the result of applied communications research, testing of new equipment, or in several cases even as the by-product of military surveillance. We concentrate here on radio wavelength observations, since they were the first to reach out beyond the traditional optical window; but there are more examples from other areas as well.

### 2.1 Communications Research:

In the course of trying to identify the source of interference to trans-Atlantic telephone communications, Karl Jansky discovered cosmic radio emission in 1933.[2] Jansky had no formal training in astronomy, and once he had determined that the interference was "of extraterrestrial origin," there was little support from the Bell Telephone Laboratory to further pin down the location in space. But Jansky learned about celestial coordinates and located the radio emission as coming from the Galactic Center. Follow-up studies by Jansky and later Grote Reber[3] showed that the Galactic radio emission, unlike all previously recognized cosmic radiation, must have a non-thermal origin. Later observations especially in Australia and the UK by former WWII radar scientists found many discrete radio sources which were later recognized as radio galaxies of unprecedented luminosity. However, following their initial identification of several strong radio sources with M87 and NGC 5128, John Bolton, Gordon Stanley, and Bruce Slee rejected the notion that they were extragalactic[4] due to the implied very large radio luminosity which they felt was unrealistic.[5] As a result of their tremendous luminosity, radio galaxies could be observed from very great distances, far beyond the reach of the best optical telescopes of the time. Radio source number counts based on early Cambridge surveys led Martin Ryle to argue in favor of an evolving universe,[6] although later observations showed that even though the Cambridge data were badly corrupted by confusion and the analysis was mathematically incorrect,[7] Ryle's conclusions were right.

Some thirty years later, in the same Bell Laboratories, Arno Penzias and Bob Wilson discovered the three degree cosmic microwave background (CMB) while trying to understand the apparent losses in a radio antenna also designed to support trans-Atlantic telephone communication, this time by satellite relay.[8] The detection of the CMB has led to a whole new industry of observations of the CMB, the rise of precision cosmology, four satellites dedicated to studying the CMB and four Nobel Prize winners.

**2.2 Military Spinoffs to Astronomy:**

In February 1942, three German warships were able to safely pass through the English Channel almost unnoticed, due to German jamming of the British radar defense. J. Stanley Hey was assigned to analyze the jamming and to develop anti-jamming techniques. A few weeks later he became aware of an apparent new form of powerful jamming of anti-aircraft radars throughout Britain. Hey recognized that the most intense noise came from the direction of the sun and that it coincided with unusually large sunspot activity.[9] He correctly concluded that the active sun was sending out intense meter wavelength radio emission. Coincidently, a few months later George Southworth, working at Bell Laboratories on the development of centimeter radar systems, independently observed intense solar radio emission from the sun. Due to military secrecy, neither Hey nor Southworth were allowed to publish their remarkable discovery. It remained for Grote Reber to be the first to report the existence of solar radio bursts, when his chart recorder went off scale while demonstrating his radio telescope to potential buyers from the US Navy.[10]

An even more dramatic and accidental astronomical discovery from a military activity came from the four Vela spacecraft which were deployed to identify γ-ray emission from possible banned Soviet testing of atomic weapons, and which instead discovered cosmic Gamma Ray Bursts.[11] GRB's are now recognized as the most powerful events in the universe and are at the core of a whole new field of research in high energy astrophysics. Subsequent purpose built spacecraft, the Compton GRO, AGILE, Integral and Fermi, have been built to study the GRB's that came to be known to the astrophysical community through their accidental detection by military space craft.

**2.3 Radio Galaxies and Quasars**:

The mystery of understanding the immense source of energy needed to power the radio galaxies was unlocked with the discovery of the very small but distant and powerful quasars. Prior to 1963, extragalactic radio sources were largely identified with elliptical galaxies or peculiar nebulae. Radio lunar occultation observations of the bright, but previously unidentified source 3C 273, showed the source to lie near a bright 13 magnitude star and a nearby "nebular wisp or jet." Based on his previous understanding of radio source identifications, Maarten

Schmidt assumed that the proper counterpart to the radio source must be the 'thin wisp,' but on a hunch he decided to first take a spectrum of the star, as it was much brighter and an easier spectroscopic target. The "star" turned out to have a redshift of 0.15, implying unprecedented luminosity from a very small volume.[12] 3C 273 is one of the brightest radio sources in the sky and is a 13$^{th}$ magnitude object optically. Much weaker radio sources had been previously routinely identified with galaxies, some as faint as 20$^{th}$ magnitude,[13] but the expectation that radio sources are associated with galaxies and not stars probably delayed for some years the identification of 3C 273 and the recognition of quasars as a major constituent of the universe. Quasars were later explained as the result of infall onto a supermassive black hole.[14] Some quarter of a century later VLBA measurements of the water maser in the nucleus of NGC 4258 gave the first direct evidence for a supermassive black hole,[15] and at the same time what is still the best direct geometric measure of the distance to a galaxy.[16] The idea of black holes had been developed much earlier by Einstein and Schwarzchild, but when he discovered quasars, Martin Schmidt wasn't looking for black holes. Indeed, the possibility that the compact radio source 3C 48 might have a large redshift was considered much earlier, but rejected due to the mental block against the large radio luminosity and small dimensions,[17,18] just as 15 years earlier Bolton, Stanley and Slee were unwilling to accept that M87 and NGC 5128 were extra galactic.[4] Schmidt and Lynden-Bell were later awarded the first Kavli Prize for the discovery and understanding of quasars.

**2.4 Interplanetary Scintillations**:

In the summer of 1962 and 1963, Cambridge University graduate student, Margaret Clark was using a radio telescope to determine accurate radio positions with the goal of identifying more quasars. But, her data for several sources proved difficult to interpret due to rapidly fluctuating signal strength especially when the sources were observed in close proximity to the sun.[19] Because her telescope had a shorter response time (time constant) than normal, she was able discern the one to two sec fluctuations that might have been smoothed over with other radio telescopes. Also, she connected the strange behavior of the scintillating sources with the unusual shape of their radio spectra which were characteristic of self-absorption, and she realized that they had to have very small angular dimensions. Despite criticism from senior associates that her equipment was faulty, she had the conviction, curiosity, and perseverance to convince others that the scintillations were real and were not due an instrumental malfunction. Tony Hewish et al.[20] later interpreted these newly discovered phenomena as due to moving inhomogeneities in the interplanetary medium named *Interplanetary Scintillations* (IPS) which provided a direct confirmation for the existence of the solar wind.

**2.5 Pulsars and Neutron Stars**:

In order to better study the structure of compact radio sources and to study the interplanetary medium, Hewish raised funds for and designed a new radio telescope with a large collecting area using an even shorter time constant to study the newly discovered IPS phenomena. Graduate student, Jocelyn Bell, was assigned to build the telescope and to write her PhD dissertation on IPS. After going through miles of chart recordings by hand, Bell noticed a strange, "scruff" on the record which repeated each day, but at the same sidereal not the same solar time. With determined curiosity, in spite of pressures from her supervisor to concentrate on her dissertation work, she soon realized that she was dealing with a previously unknown phenomenon, radio sources that pulsed with periods of the order of one second, and later named *pulsars.*[21] After dismissing an interpretation in terms of "Little Green Men," pulsars were soon understood to be rapidly rotating neutron stars.[22] The possible existence of neutron stars had been discussed much earlier, only a year after the discovery of neutrons by Walter Baade and Fritz Zwicky,[23] but this paper was unknown to Bell and Hewish, and it played no role in the discovery of pulsars. "For his decisive role in the discovery of pulsars," Tony Hewish shared the 1974 Nobel Prize with Martin Ryle.

As it later turned out, around the same time, Air Force Officer Charles Schisler[24] had independently discovered ten pulsars during a tour of duty in Alaska at the Ballistic Missile Early Warning Site. As a result of his understanding of celestial navigation previously obtained as a bomber pilot, he recognized that a signal that reappeared every 23hr and 56min was extraterrestrial and not from a Soviet ICBM. On his own, Schisler followed up his discovery at the Fairbanks library and recognized that one of his signals came from the Crab nebula. Only after the recent deactivation of the radar system was this work declassified and released to the public.

**2.6 Gravitational Radiation***:*

Following the discovery of pulsars, many astronomers set out to make accurate timing measurements in order to better understand their energetics, spin-down rates, etc. From careful timing measurements at Arecibo, Russ Hulse and Joe Taylor realized that the pulsar PSR 1913+16, was part of a binary system, and they predicted that the orbit would decay due to the effects of gravitational radiation.[25] Hulse and Taylor later shared the 1993 Nobel Prize for their role finding the first experimental evidence for gravitational radiation.

**2.7 Extra-solar Planets**:

Precision timing observations led Alex Wolszczan and Dale Frail,[26] to realize that small perturbations in the pulse arrival times from the millisecond pulsar PSR 1257+12 were due to at least two planet-sized bodies orbiting the pulsar. Although followed by many other detections of extra solar planets, for many years PSR 1257+12 remained the only known extra solar planetary system, and the only earth-sized planets known.

**2.8 Ignored Predictions**:

Although the existence of the CMB was predicted, and even earlier observed, but unrecognized, the theoretical prediction played no role in the discovery by Arno Penzias and Robert Wilson. As is well known, Penzias and Wilson were trying to find the source of noise in the 20-m horn parabola which was built as the ground link for the Echo balloon relay satellite. After painstaking troubleshooting and eliminating all possible instrumental sources, they had concluded that their excess noise could not be understood in terms of any instrumental effect. Meanwhile, not far away in Princeton, Robert Dicke and his colleagues were building a radiometer to follow up on Dicke's prediction that it might be possible to detect the remnants of the Big Bang. They were beaten out by Penzias and Wilson for the Nobel Prize, although all that Penzias and Wilson were trying to do was to understand their antenna. The existence of a cosmic microwave background (CMB) had been predicted by George Gamow 20 years earlier,[27] but this was unknown to either Penzias and Wilkinson or to Dicke, and played no role in what was certainly the most important discovery in cosmology since Hubble's discovery of the expanding universe.

In fact, as Dave Wilkinson[28] has commented, using a simple system he had built to measure atmospheric water vapor, Dicke could have detected the CMB back in 1946 near the time of Gamow's prediction. By 1965, everyone had forgotten Gamow's prediction. Everyone, that is, except the Russian scientists, A. G. Doroshkevich and Igor Novikov,[29] who were more familiar than the Americans with a 1961 Bell Labs paper by E. A. Ohm[30] that reported an excess antenna temperature. The Russians were looking for experimental evidence of what they called "the relict radiation," but they mistranslated Ohm's paper and incorrectly concluded that the excess temperature observed by Ohm was due to atmospheric radiation. As it later turned out, the CMB had been detected much earlier by Andrew McKellar who noted that interstellar CN had an excitation temperature of 2.3 K. No process was then known to produce this level of excitation[31], and although this was a long standing puzzle in astrophysics, no one made the connection with Gamow's prediction until after the Bell Labs detection.

**2.9 Beware of Theoreticians**:

In 1968, a proposal to NRAO to search for $H_2O$ emission with the 140-ft radio telescope was rejected because theoretical arguments suggested that the water molecule would be too weak to detect. However, subsequent observations by Cheung et al.[32] with only a 6-m antenna observed remarkably strong $H_2O$ due to maser action. With hindsight, $H_2O$ masers could probably also have been detected even before the HI line with the simple 1.3 cm radiometer and 18 inch dish used more than 20 years earlier by Dicke and Beringer[33] to measure atmospheric water vapor.

**2.10 Close to Home in the Solar System**:

Even within the solar system, there have been many surprises. Stefan's Law predicts the expected surface temperature of each planet depending only on the solar constant, the distance from the sun and the albedo. Passive radio studies simply intended to measure the thermal emission and surface temperature from each planet turned up surprises with every planet except Mars.

Ever since Giovanni Schiaparelli thought he repeatedly saw the same markings on the surface of Mercury, it was widely accepted that Mercury rotated every 88 days in synchronism with its orbital motion, and so it was expected that the daytime side must be very hot, and the eternally unheated night side incredibly cold. But 10 cm radio measurements showed both the day and night side to be close to room temperature, although this was misinterpreted by the author as convection from via a thin atmosphere.[34] Subsequent radar observations by Gordon Pettengill and Rolf Dyce showed directly[35] that Mercury rotates with a 59 day period in 2/3 synchronism with the revolution. So for every two revolutions of Mercury around the sun, there are three full rotations about its axis, and thus for every other perihelion passage the same face is visible from the Earth; and so for more than a hundred years astronomers had apparently ignored half of their admittedly difficult observations of the sparse surface markings. Retroactive "*predictions"* quickly showed, in fact, that an 88 day period would not be stable, and that a 59 day period is the result of Mercury's very eccentric orbit.[36]

In the late 1950's and early 1960's Russian, British, and American radar scientists were competing to be the first to detect radar echoes from Venus. There was no particular scientific motivation, other than to be first, and to demonstrate the effectiveness of their sensitive receivers, powerful transmitters, newly devised high speed digital recording and sophisticated signal analysis techniques. However, the echoes from Venus showed that it unexpectedly rotated in the retrograde direction and gave a new value for its distance and thus the AU, which was more accurate by about a factor of 100 than the previously accepted value.[37] Passive radio observations showed that the surface of Venus was surprisingly hot, near 600 C.[38] This was later explained as due to a greenhouse effect, a phenomenon subsequently applied to global warming on the Earth.

Unrealistically high temperatures were also measured for Jupiter, but the apparent temperature increased with wavelength, suggesting a non thermal origin.[39] Speculation that the non thermal radiation from Jupiter might be due to a powerful analogue of the Earth's Van Allen Belts, was later confirmed with direct radio interferometric imaging of Jupiter's radiation belts.[40] Even earlier, Bernie Burke and Ken Franklin[41] had detected intense decametric busts from Jupiter while setting up their new antenna to observe the Crab nebula which fortuitously happened to be close to the same declination as Jupiter and passed through their beam every night. Multi-wavelength radiometric observations of Saturn, Uranus, and Neptune, later indicated temperatures well in excess of that expected from heating by the sun, giving the first suggestions of an internal source of heat due to radioactive decay.

## 3. The Lessons of History: The Design of New Facilities

Theoretical calculations can be dangerous in the planning and design of new instruments as well as for inhibiting new discoveries. The Jodrell Bank 250-ft radio telescope was designed to detect radar echoes from cosmic ray air showers which P. M. S. Blackett and A. C. B. Lovell[42] calculated would be possible with a large antenna. Although it was pointed out to Lovell that recombination in the ionized cosmic ray trail would greatly suppress the echo below detectability, Lovell claimed to have forgotten or not paid attention to the correct calculations and built the 250-ft reflector anyway.[43]

The Arecibo 1000-ft dish was designed by Bill Gordon in the 1950's for ionospheric backscatter experiments, not for radio astronomy. However, it later became apparent that Gordon had overestimated the spectral width of the returned echoes in calculating the dish size needed to detect echoes from the ionosphere, and that a much smaller (and very much cheaper) dish would be sufficient for the intended ionosphere experiments. However, by then enthusiasm for a 1000-ft dish had grown, and Gordon was able to obtain construction funds from the military who were obsessed with anything that they might learn about the ionosphere in order to perhaps detect a signature of incoming Russian missiles.[44] The 1000-ft Arecibo telescope was built as designed and still has, by far, the largest collecting area of any radio telescope and is the model for a yet larger similar facility being built in China.

Interestingly, although the scientific justification leading to the funding of Jodrell Bank and Arecibo telescopes were not correct, based on these wrong arguments, the telescopes were built, and they have had nearly a 50 year record of successes in ways that the original advocates could not have possibly imagined, including the detection of the effects of gravitational radiation, the discovery of the first extra-solar planetary system, the surprising measurement of the rotation period of Mercury, the first convincing evidence for large scale extra galactic structures in the universe[45] and the return of the first photographs from the far side of the moon. What if Lovell and Gordon had correctly understood the true sensitivity needed for their intended meteor and ionosphere experiments?

## 4. The Human Factors

Three transformational astronomical discoveries, each of which defined new fields for research, Jansky's detection of extraterrestrial radio emission, the detection of strong non thermal radio emission from the sun by Southworth, and the detection of the cosmic microwave background by Penzias and Wilson, all occurred at the same Bell Telephone Laboratories. The first two of these Bell Labs discoveries, like the Los Alamos discovery of GRBs, were not made by people trained in astronomy, or even as a result of a basic research program. Although they were quite independent, it is perhaps no coincidence that the Bell Telephone Laboratories, with its rich heritage of independent research and concentration of scientists and engineers such as Edmond Bruce, George Southworth, Claude Shannon, Charles Townes, Harry Nyquist, William Shockley, John Bardeen, Walter Brattain, and Philip Smith (of Smith chart fame) with their wide range of expertise provided a fertile ground for exploration and discovery.

As Hey later remarked,[9] *The previous failure of other workers to recognize abnormal solar radio emission illustrates, I think, the stultifying effect of clinging to established* *viewpoints, in this case biased by early negative attempts. The intense noise is so strong that it had almost clamoured to be observed in the past.* A similar remark could be made about water vapor masers, pulsars, or the CMB. Arguably, "luck," plays as much a role in scientific discovery, as careful planning. But, as wisely commented by Louis Pasteur, *"In the field of observation, chance favors the prepared mind,"* or from Gary Player's approach to golf, "*The harder I practice the luckier I get."* Scientific discoveries come from the right person, in the right place, at the right time and doing the right thing using the right instruments.

The impact of the next generation of astronomical facilities will not only depend on the cleverness of the scientists who use them, but on the cleverness of their designers to obtain better sensitivity, image quality, resolution, field of view, time domain coverage, or the opportunity to explore new parts of electromagnetic and non electromagnetic spectrum (e.g., gravity). Equally important, will be the training of the next generation of scientists so that they understand the instruments they use. So, like Jansky, Reber, Southworth, Clarke, Bell, Schisler, Penzias, and Wilson, they are able to explore unanticipated results for more than their immediate intended purpose.

## 5. New Discoveries

Martin Harwit[46,47] has discussed the historical role of discovery in astronomy from antiquity to the present, and in particular the explosive growth starting early in the 20$^{th}$ century with the construction of a new generation of powerful observational facilities. Harwit introduced "discovery" as a topic for rational debate within the astronomical community — recognizing that new phenomena can be systematically unearthed by the exploration of new areas of parameter space by the application of new technology.

Most of the phenomena studied by modern telescopes were unknown even 50 years ago, and many were discovered from observations made at radio wavelengths by using increasingly more powerful instruments, often motivated by solving other problems, or by scientists just following their curiosity. The discovery of new phenomena has been, and will continue to be more transformational than the explanation of old questions posed by previous discoveries. The history of astronomy suggests that the opportunities for the discovery of new phenomena are optimized when new facilities have at least an order of magnitude improvement in capability in sensitivity, resolution or image quality, temporal extent and resolution, or spectral coverage and resolution. But, it will be equally important that scientists, especially students and young scientists, understand their instruments and their data, and that they are given the opportunity to follow their curiosity.

While the potential for new astronomical discoveries will be heavily dependent on the application of innovative new technologies to the next generation of astronomical instruments, a lot will also depend on the quality of the scientists with good understanding of their instrument and an ability and willingness to explore and accept new ideas and not to sweep seemingly anomalous results under the rug as due to "instrumental effects." It will be equally important for those who are in a position to filter research ideas, either as grant or observing time referees, as managers of facilities, or as mentors to young scientists, not to dismiss as "butterfly collecting," proposed investigations which explore new areas of phase space *without having predefined the result they are looking for.*

While it is fashionable to consider that research follows the textbook picture whereby theories are first formulated and then followed by experimental or observational tests, the converse is often true. Progress must also allow for new discoveries, as well as for the explanation of old discoveries. These remarks are not original, and many scientists lament the apparent conservatism of funding agencies and governments in funding research grants and new facilities who seemingly want to know what will be discovered as a result of proposed research.

However, governments and funding agencies don't act on their own but depend on peer review. The problem is us!

We are grateful to Marshall, Cohen, Bob Wilson, Maarten Schmidt, Dale Frail, Guenther Elste, and Bill Howard who have helped to clarify some historical points. The National Radio Astronomy Observatory is operated by Associated Universities Inc. under Cooperative Agreement with the National Science Foundation. Basic research in radio astronomy at the NRL is supported by 6.1 base funding.

[18] J. L. Greenstein, 1963, unpublished manuscript.

[19] M. Clark, 1964, *PhD Thesis*, Cambridge University.

[20] A. Hewish, P. F. Scott, and D. Wills, 1964, *Interplanetary Scintillation of Small Diameter Radio Sources,* Nature, 203, 1214.

[21] A. Hewish et al, 1968, *Observation of a Rapidly Pulsating Radio Source*, Nature 217, 709.

[22] T. Gold, 1968, *Rotating Neutron Stars as the Origin of the Pulsating Radio Sources,* Nature 218, 73.

[23] W. Baade and F. Zwicky, 1933, Phys Rev, 46, 76.

[24] C. Schisler, *An Independent 1967 Discovery of Pulsars*, in *40 years of Pulsars*, p. 642, AIP, 2008.

[25] R. A. Hulse and J. H. Taylor, 1975, *Discovery of a Pulsar in a Binary System,* ApJ, 195, L51.

[26] A. Wolszczan, and D. A. Frail, 1992, *A Planetary System around the Millisecond Pulsar PSR1257 + 12,* Nature, 355, 145.

[27] G. Gamov, 1946, Phys. Rev. 70, 572.

[28] D. Wilkinson, 1983,, in *Serendipitous Discoveries in Radio Astronomy,* p. 176.

[29] A. G. Doroshkevich and I. Novikov, 1964, Sov. Physics Doklady, 9, 111.

[30] E. A. Ohm, 1961, Bell System Technical Journal, 40, 1065.

[31] A. McKellar, 1941, *Molecular Lines from the Lowest States of Diatomic Molecules Composed of Atoms Probably Present in Interstellar Space*, Publ. Dom. Astrophys. Obs. 7, 251.

[32] A. C. Cheung et al., 1969, *Detection of Water in Interstellar Regions by its Microwave Radiation,* Nature 221, 626.

[33] R. H. Dicke and R. Beringer, 1946, *Microwave Radiation from the Sun and Moon*, ApJ, 103, 375.

[34] K. I. Kellermann, 1965, *11-cm Observations of the Temperature of Mercury,* Nature 205, 10.

[35] G. Pettengill and R. Dyce, 1965, *A Radar Determination of the Rotation of the Planet Mercury,* Nature 206, 1240.

[36] S. J. Peale and T. Gold, 1965, *Rotation of the Planet Mercury*, Nature 203, 1241.

[37] G. Pettengill and R Price, 1961, *Radar echoes from Venus and a new determination of the solar parallax,* Planetary and Space Science, 5 (1), 71.

[38] C. H. Mayer, T. P. McCullough, R. M. Sloanaker, 1958, *Observations of Venus at 3.15-CM Wave Length,* ApJ, 127, 1.

[39] F. D. Drake and H. Hvatum, 1959, *Non-thermal Microwave Radiation from Jupiter,* AJ, 64, 329.

[40] V. Radhakrishnan and J. Roberts, 1960, *Polarization and Angular Extent of the 960-Megacycle Radiation from Jupiter,* AJ, 65, 498.